\documentclass{article}
\usepackage[utf8]{inputenc}
\usepackage[a4paper, total={5.5in, 9.7in}]{geometry}

\usepackage{multicol}
\usepackage{authblk} % for author affiliations
\usepackage{graphicx} % for inserting figures
\usepackage{physics}
\usepackage{amsmath}
\usepackage{mathtools}
\usepackage{verbatim} % package for using the "comment" syntax
\usepackage{float} % for using [H] to place the figure exactly at this position
\usepackage{xcolor} % for coloring text
\usepackage{cite} % for citing to be shown as a range like [1-5]
\usepackage[resetlabels,labeled]{multibib}

\newcommand{\supplementaryinformation}{%
  \setcounter{section}{0}% Reset section counter
  \let\oldthesection\thesection% Capture section numbering scheme
  \renewcommand{\thesection}{S\oldthesection}% Prefix section number with S
  
  \setcounter{figure}{0}% Reset figure counter
  \let\oldthefigure\thefigure% Capture figure numbering scheme
  \renewcommand{\thefigure}{S\oldthefigure}% Prefix figure number with S
  
  \setcounter{equation}{0}% Reset equation counter
  \let\oldtheequation\theequation% Capture equation numbering scheme
  \renewcommand{\theequation}{S\oldtheequation}% Prefix equation number with S
  
  \part*{Supplementary Information}% Set supplementary information
} %

\title{Thickness mapping and layer number identification of exfoliated van der Waals materials by Fourier imaging micro-ellipsometry}
\author[1]{Ralfy Kenaz}
\author[1]{Saptarshi Ghosh}
\author[1]{Pradheesh Ramachandran}
\author[2]{Kenji Watanabe}
\author[3]{Takashi Taniguchi}
\author[1]{Hadar Steinberg}
\author[1,*]{Ronen Rapaport}
\affil[1]{\small{Racah Institute of Physics, The Hebrew University of Jerusalem, Jerusalem 9190401, Israel}}
\affil[2]{\small{Research Center for Functional Materials, National Institute for Materials Science, 1-1 Namiki, Tsukuba 305-0044, Japan}}
\affil[3]{\small{International Center for Materials Nanoarchitectonics, National Institute for Materials Science, 1-1 Namiki, Tsukuba 305-0044, Japan}}

\affil[*]{\small{ronen.rapaport@huji.ac.il}}

\date{}

\begin{document}

\maketitle

%TC:ignore
\begin{abstract}

As properties of mono- to few layers of exfoliated van der Waals heterostructures are heavily dependent on their thicknesses, accurate thickness measurement becomes imperative in their study. Commonly used atomic force microscopy and Raman spectroscopy techniques may be invasive and produce inconclusive results. Alternatively, spectroscopic ellipsometry is limited by tens-of-microns lateral resolution and/or low data acquisition rates, inhibiting its utilization for micro-scale exfoliated flakes. In this work, we demonstrate a Fourier imaging spectroscopic micro-ellipsometer with sub-5 microns lateral resolution along with fast data acquisition rate and present angstrom-level accurate and consistent thickness mapping on mono-, bi- and trilayers of graphene, hexagonal boron nitride and transition metal dichalcogenide (MoS$_2$, WS$_2$, MoSe$_2$, WSe$_2$) flakes. We show that the optical microscope integrated ellipsometer can also map minute thickness variations over a micro-scale flake. In addition, our system addresses the pertinent issue of identifying monolayer thick hBN.

\end{abstract}
%TC:endignore

\section{Introduction}

Ever since mechanical exfoliation was first used for the isolation of single-layer graphene from graphite \cite{Novoselov2004ElectricFilms}, a range of exfoliable materials including hexagonal boron nitride (hBN) and transition metal dichalcogenides (TMDs) with strong in-plane and weak van der Waals (vdW) out-of-plane molecular bonds have been appended to the 2D inventory. Either individually or in stacks, these materials are constantly charting new frontiers both in application oriented research such as electrocatalysis and renewable energy \cite{Chandrasekaran2020ElectronicReactions}, nanophotonics \cite{Zotev2022TransitionInteractions, Zheng2018LightApplications}, quantum optics \cite{Wrachtrup2015SingleTemperature, Tran2015QuantumMonolayers, Palacios-Berraquero2016AtomicallyDiodes}, and in fundamental science such as exciton physics\cite{Srivastava2015ValleyWSe2, Zhang2019HighlyHeterostructure, Regan2022EmergingHeterobilayers}, spintronics and valleytronics \cite{Xu2014SpinDichalcogenides, Arora2016ValleyFields}.

Thickness (and thus layer number) is a crucial parameter in many exotic effects involving vdW layered structures, as was shown in superconducting TMDs \cite{ Costanzo2016Gate-inducedCrystals}, low angle twisted systems \cite{Cao2018UnconventionalSuperlattices} and freestanding planar waveguides for propagating exciton-polariton systems \cite{Hu2017ImagingWaveguides}. Its effect is equally important in bandgap analytics of 2D systems \cite{Lezama2015Indirect-to-DirectMoTe2}, stacked TMDs realizing tunable Hamiltonian models \cite{Xu2022AWSe2, Gotting2022Moire-Bose-HubbardHeterostructures}, generalized Wigner crystals \cite{Zhou2021BilayerHeterostructure} and exotic correlated exciton states\cite{Slobodkin2020QuantumHeterostructures, Zimmerman2022CollectiveExcitons, Shimazaki2021OpticalState}.

Similarly, in light harvesting it is now established that TMD based solar cells require sub-1 nm thick flakes as their absorption decreases with increasing thickness \cite{Bernardi2013ExtraordinaryMaterials}. Thickness determination is also important when considering the role of TMDs and hBN as tunnel barriers for charged carriers in tunneling based devices, as the tunneling rate is suppressed exponentially with the number of layers \cite{Britnell2012ElectronBarriers, Dvir2018TunnelingJunctions}. Thickness dictates the coupling of single defects in electrically gated thin vdW devices \cite{Keren2020Quantum-dotGraphene}. Atomic level thickness is also imperative for TMD based single photon emitters \cite{Parto2021Defect150K, He2022RoomTransfer}.

Despite layer number and thickness being crucial for so many applications, identifying layer numbers of exfoliated flakes and accurately measuring the thickness and thickness variations by a simple technique remains challenging.

A number of factors impede precise thickness determination of exfoliated flakes. Effective flake thickness may depend on conditions like conformation and ambient conditions like temperature. Uniaxial strain and pressure have both been reported to alter the bond length and the consequent thickness of monolayers \cite{Chu2016StructuralMoS2}. Furthermore, presence of physisorbed organic molecules on the surface increases their measured thicknesses \cite{Falin2021MechanicalWTe2}. Substrate-TMD interfaces are also prone to the presence of air gaps which are otherwise undetectable by standard characterization techniques, resulting in erroneous atomic force microscopy (AFM) line profiles and significant deterioration of the device performance \cite{Chiu2018SynthesisMultijunctions}. Thus accurate thickness estimation of exfoliated materials by a non-invasive, fast and repeatable method is of prime importance. 

Optical microscopy can be used for estimating the number of layers of such exfoliated flakes based on the color contrast or by fitting the optical contrast spectra by a Fresnel law based model \cite{Zhao2020ThicknessMethods}. Notably, optical microscope imaging of photoluminescence (PL) spectra could successfully discern the interlayer coupling between hetero-bilayers and consequently distinguish between number of layers as well as twist angle between them \cite{Alexeev2017ImagingMicroscope}. However, this method is appropriate for estimating the number of layers and not the actual thickness. Moreover, optical microscopy involves uncertainties due to influence from the microscope hardware, illumination spectra, imaging conditions and oxide thickness of the substrate \cite{Gorbachev2011HuntingSignatures, Golla2013OpticalFlakes}. Furthermore, not all vdW materials have sufficient monolayer PL at the visible optical wavelength range. While TMDs and graphene can be distinguished by contrast on a SiO$_2$/Si substrate, the visible- and near-infrared transparency of mono- and few-layer hBN make their identification a challenging proposition (a maximum of 2.5\% contrast is achievable for monolayer hBN) \cite{Crovetto2018NondestructiveMonolayer , Gorbachev2011HuntingSignatures}.

Raman spectroscopy and AFM are accepted means of estimating the layer numbers and thickness of vdW structures. While AFM is relatively more complex, Raman spectroscopy exposes the material to high laser powers which might result in localized stress due to thermal expansion mismatch \cite{Ferrari2013RamanGraphene, Wang2022CriticalNanomaterials}, in addition to weak signal response \cite{Malard2021StudyingTERS}. Moreover, Raman analysis might be inconclusive for distinguishing among different layers for certain TMDs \cite{Terrones2014NewDichalcogenides}. In WSe$_2$, a shift of 1 cm$^{-1}$ in the secondary peak is used to distinguish between bilayer and trilayer \cite{Sahin2013AnomalousWSe2}, thus offering small tolerance and require very high signal to noise ratios. For AFM, differences in gradients of the attractive forces and lateral forces on the material and the substrate \cite{Li2016LayerOxidation}, presence of surface adsorbents \cite{Li2012FromScattering} and anomalies due to tip-sample interactions \cite{Nemes-Incze2008AnomaliesMicroscopy} are common causes for misinterpreting the thickness values. Contact-mode AFM is also moderately invasive that might damage the investigated flake and is prone to faulty thickness measurement owing to small spring constant of the probe \cite{Rapuc2021NanotribologyMonolayers}. Moreover, due to the undulation on a silicon oxide substrate, AFM scan steps from the substrate to the flake are less reliable than steps across boundaries within a flake. 

Alternately, spectroscopic ellipsometry provides a non-invasive yet accurate way for optical constants and thickness measurements of two-dimensional vdW materials \cite{Wurstbauer2010ImagingGraphene, Funke2017Spectroscopic2D-materials, Ermolaev2020BroadbandMoS2}. The technique is highly sensitive and was successfully used in measuring the thickness of hBN monolayers with angstrom-level precision \cite{Crovetto2018NondestructiveMonolayer}. However, current ellipsometers with integrated focusing optics cannot resolve a spot-size smaller than $\sim$50 microns at most \cite{Kravets2019MeasurementsModulators}, \textit{making them incapable of addressing exfoliated flakes of vdW materials which are commonly smaller in lateral dimensions}. Another class of commercial spectroscopic ellipsometers utilize imaging ellipsometry with a high lateral resolution to address micron-scale flakes. However, they use a monochromator (or spectral filters) which accounts for a single-wavelength at-a-time, often resulting in very long data acquisition times for spectrally resolved information \cite{Park2018ComparisonRotator}. Thus, practical use of current ellipsometers for exfoliated vdW materials is limited, together with the fact that these ellipsometers are stand-alone tools that cannot be easily integrated with typical optical experimental setups.

In this paper, we demonstrate an accurate way for thickness measurement of vdW materials with angstrom-level accuracy by our recently developed Fourier imaging spectroscopic micro-ellipsometer (SME) that can be integrated into any optical microscope\cite{Kenaz2022MappingResolution}. Our ellipsometer has a sub-5 microns lateral resolution (one order-of-magnitude higher compared to focused-beam spectroscopic ellipsometry \cite{Meshkova2018TheFilms, Kravets2019MeasurementsModulators}). Additionally, with a fast data acquisition rate of a few seconds, ellipsometric data with fine spectral and angular resolution can be recorded per lateral position. Thus it is capable of performing thickness measurements and mapping the thickness variations of exfoliated micro-scale vdW flakes.

We show that the SME can accurately measure and map the thickness of exfoliated flakes of vdW materials and can easily distinguish among monolayers, bilayers and trilayers of different genres of vdW flakes including conductive graphene, various type-II semiconductor TMDs and wide band gap dielectric hBN. These materials are chosen as they are the constituents in various heterostructure devices being investigated by the 2D research community.

\begin{figure}[t]
\centering
\includegraphics[width=1\textwidth]{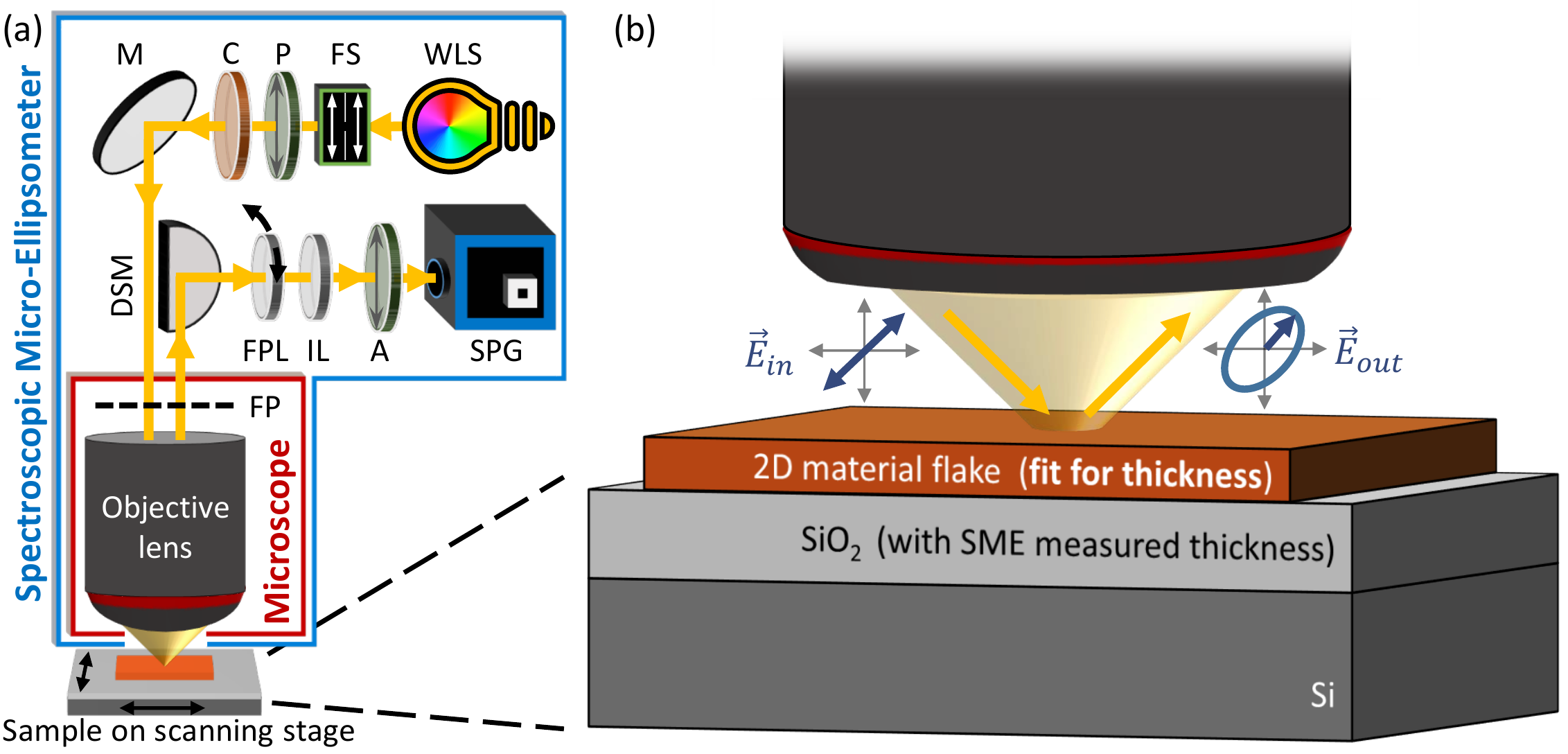}
     \caption{(a) Schematic of the spectroscopic micro-ellipsometer (SME) introduced in reference \cite{Kenaz2022MappingResolution}. WLS: White light source, FS: Field stop, P: Polarizer, C: Compensator (Quarter-wave retarder), M: Mirror, DSM: D-shaped mirror, FP: Fourier plane, FPL: Fourier plane lens (on a flip mount - for interchanging between sample-imaging microscope mode and the Fourier plane imaging ellipsometry mode), IL: Imaging lens, A: Analyzer (polarizer), SPG: Spectrograph with a detector array. (b) Multiple-angle reflection of white light from the sample causes sample-dependent variations per incident angle on the polarization state, namely, transformation of linear polarization ($\vec{E}^{\,}_{in}$) to elliptical polarization ($\vec{E}^{\,}_{out}$). Previously measured nearby SiO$_2$ thickness and the optical constants from the literature for the measured 2D material are used in the model for extraction of the flake thickness.}
\label{fig:measurement}
\end{figure}

\section{Measurement, modelling and fitting}

To show the wide range of capabilities of the SME, various vdW materials were tape exfoliated and transferred by polydimethylsiloxane (PDMS) assisted dry transfer method onto silicon chips with a 285 nm SiO$_2$ (P-type $<$100$>$ prime grade silicon wafers from NOVA Wafers with a thermal oxide thickness of 2850 \AA) \cite{Benameur2011VisibilityNanolayers}. Based on the contrast under an optical microscope, possible candidates for mono, bi- and trilayers were identified for graphene, hBN and TMDs (MoS$_2$, WS$_2$, MoSe$_2$ and WSe$_2$), to be eventually measured with the SME for their thicknesses. Finding candidates for monolayer hBN in optical microscope was extremely challenging and required multiple iterations. The investigated flakes were pre-annealed in forming gas before measurement to remove surface adsorbents as well as the entrapped water molecules between the flake and the substrate.

Figure \ref{fig:measurement} shows the schematic design of the SME and the illustration for the flake measurement. After locating the flake of interest under the objective lens (NA = 0.9) of the SME in microscope mode (see Figure \ref{fig:measurement}(a)), the SME is switched to ellipsometry measurement mode. Measurements were performed first on the substrate just outside the periphery of the flake and subsequently on the flake, obtaining the spectrally and angularly resolved spectroscopic ellipsometry data of both points (or areas in case of mapping experiments). At each measurement point, the SME took four consecutive first-order images of the objective lens Fourier (back focal) plane at different polarization settings, providing spectrally and angularly resolved reflection intensity information, which is then processed to calculate the ellipsometric data of the area. Our work on development of the SME \cite{Kenaz2022MappingResolution} gives a detailed discussion on its operation principle, the data acquisition method, and the instrument performance. The data obtained from the substrate is modelled as Air/SiO$_2$/Si layered structure and fitted for the oxide thickness to obtain its exact value. Then the flake data is modelled as Air/Flake/SiO$_2$/Si layered structure and the previously measured SiO$_2$ thickness value is used in the model. The thickness of the oxide layer in the vicinity of the flake is assumed to not fluctuate considerably under the flake. The thin film thickness measurement accuracy of the SME was reported to be in good agreement with a commercial ellipsometer in our previous work \cite{Kenaz2022MappingResolution}. Next, depending on the flake material, the complex refractive index values obtained from references \cite{Weber2010OpticalEllipsometry,Hsu2019Thickness-DependentWSe2, Adachi1999Hexagonalh-BN} are used in the model and the thickness of the flake is fitted for. The obtained thickness value of the flake is used to determine the number of layers. Due to different experimental methods used in the literature to extract the optical constants, they might not exactly coincide with those of the flakes measured by the SME, however, these possible deviations in optical constants do not interfere with the ability of the model to predict the number of layers of the measured flakes. 

For modelling and fitting, WVASE\textsuperscript{\tiny\textregistered} and CompleteEASE\textsuperscript{\tiny\textregistered} ellipsometry data analysis software (J.A. Woollam Co., Inc.) are used.

\begin{figure}[!ht]
\centering
\includegraphics[width=\textwidth]{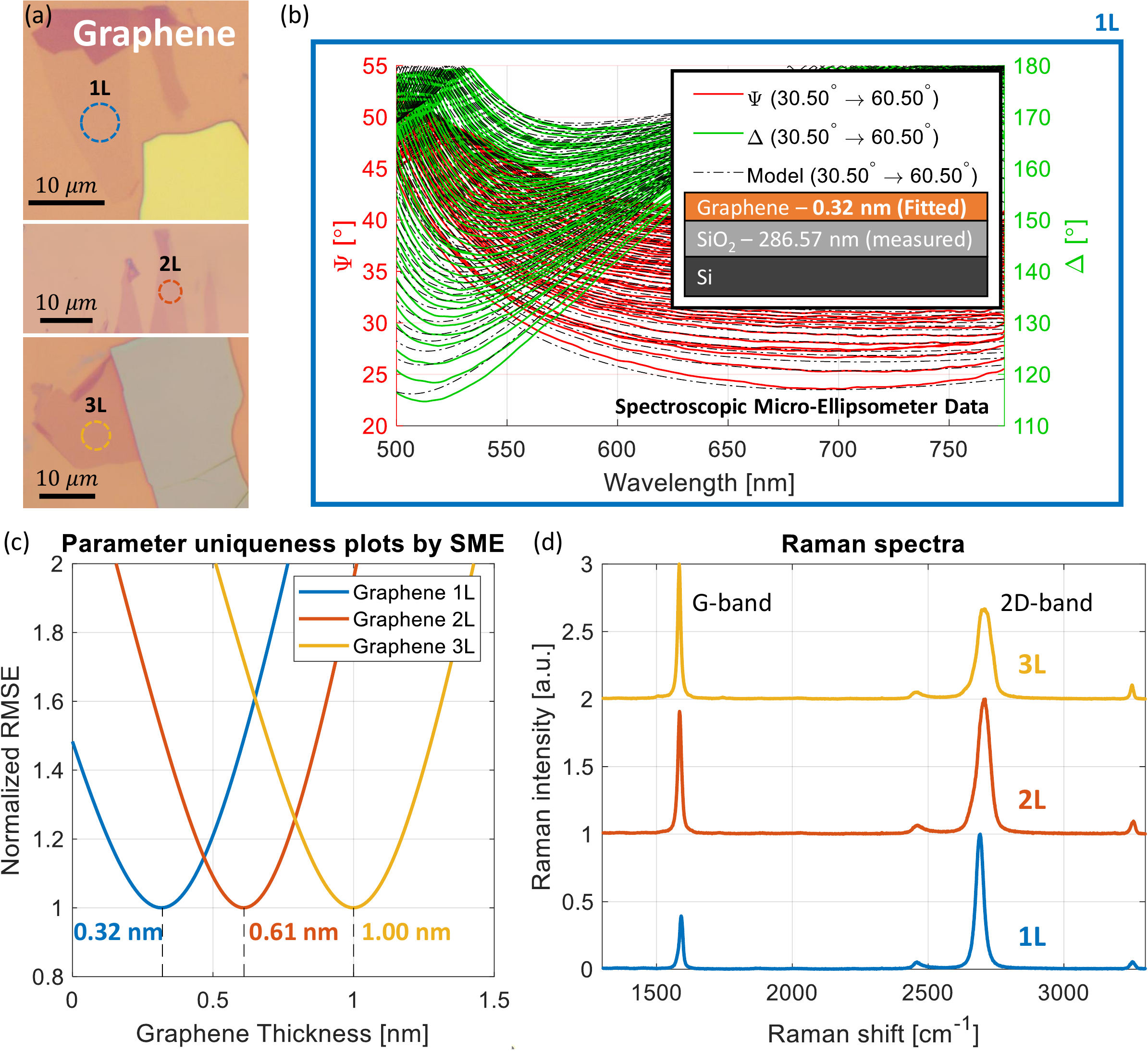}
     \caption{(a) Optical microscope images of monolayer (1L), bilayer (2L) and trilayer (3L) graphene flakes on 285 nm SiO$_2$/Si with illustrated spectroscopic micro-ellipsometer (SME) measurement spots having a diameter of 5 $\mu m$. (b) Single-measurement SME data on the monolayer consisting of spectrally and angularly resolved $\Psi$ and $\Delta$ values. The model is illustrated in the plot legend, with inputs of SME measured SiO$_2$ thickness and graphene optical constants \cite{Weber2010OpticalEllipsometry}, resulting in best fit value at thickness of 0.32 nm. The same procedure is repeated for bilayer and trilayer graphene. (c) The parameter uniqueness plots by the SME for monolayer, bilayer and trilayer measurements point to graphene thicknesses of 0.32 nm, 0.61 nm and 1.00 nm, respectively. (d) The Raman spectra measured on the same flakes confirm their mono-, bi- and trilayer nature measured by the SME \cite{Li2009Large-areaFoils, Huang2020Large-areaFoil}.}
\label{fig:graphene_example}
\end{figure}

\section{Results}

The optical microscope images of monolayer (1L), bilayer (2L) and trilayer (3L) graphene with illustrated 5 $\mu m$ diameter SME measurement spots are shown in Figure \ref{fig:graphene_example}(a) (these flakes were also used in our previous work \cite{Kenaz2022MappingResolution}). Figure \ref{fig:graphene_example}(b) plots the SME data of one measurement from the monolayer graphene, consisting ellipsometric parameters $\Psi$ and $\Delta$ at 551 wavelength points between 500 nm and 775 nm, and at 52 different values of angles-of-incidence between 30.50$^{\circ}$ and 60.50$^{\circ}$. The change in light polarization reflected from the sample is represented by parameters $\Psi$ and $\Delta$ ($\Psi$ is related to the amplitude ratio between the s- and p-components of the polarized light, whereas $\Delta$ is the phase difference between them, see Ref. \cite{Kenaz2022MappingResolution} for more details). Importantly, this whole set of data is acquired in just 4 exposures (in different measurement polarization settings), with a total measurement time of $\sim$10 seconds.

The measured oxide thickness and the complex refractive index of graphene obtained from Ref. \cite{Weber2010OpticalEllipsometry} are used in the model to fit for the thickness, as shown in Figure \ref{fig:graphene_example}(b) for the monolayer graphene. The same procedure is repeated for bilayer and trilayer graphene flakes. Figure \ref{fig:graphene_example}(c) shows the parameter uniqueness plots for all three measurements, normalized to their corresponding minimum values. These plots represent the relative error between the data and the model fit as a function of the thickness value. The global minimum of each curve provides the best fit between the ellipsometric data (i.e., $\Psi$ and $\Delta$) and the model, which for the monolayer occurs at graphene thickness of 0.32 nm. This is in good agreement with the theoretical thickness of 0.34 nm for single-layer graphene \cite{Liu2021ControllingMembranes}. Similarly, thicknesses of 0.61 nm and 1.00 nm are obtained for a bilayer and a trilayer graphene respectively, again in agreement with the literature \cite{Nemes-Incze2008AnomaliesMicroscopy}. Each measurement on mono-, bi- and trilayer graphene is repeated 10 times to demonstrate instrumental accuracy in thickness results. Standard deviations of $\sim$0.02 nm is obtained for all the three sets of measurements. Finally, the Raman spectra of the same flakes are measured, as shown in Figure \ref{fig:graphene_example}(d), normalized and vertically displaced for better visibility, confirming the findings of the SME. The peak intensity ratio of the 2D-band to the G-band (I$_{2D}$/I$_{G}$ $\sim$2.5) and the symmetric 2D-band at $\omega$ $\sim$2690 cm$^{-1}$ with a full width at half-maximum (FWHM) $\sim$33 cm$^{-1}$ provide an exclusive signature for monolayer graphene \cite{Li2009Large-areaFoils, Wu2011ControlElectrodes, Huang2020Large-areaFoil}. Similarly, I$_{2D}$/I$_{G}$ $\sim$1.1, 0.67 intensity ratios and asymmetric 2D-bands with FWHM $\sim$53, 62 cm$^{-1}$ show the typical features of bilayer \cite{Wu2011ControlElectrodes, Huang2020Large-areaFoil, Delikoukos2020Doping-InducedManipulation} and trilayer graphene \cite{Wu2011ControlElectrodes, Huang2020Large-areaFoil, Delikoukos2020Doping-InducedManipulation} respectively.

\begin{figure}[!h]
\centering
\includegraphics[width=1\textwidth]{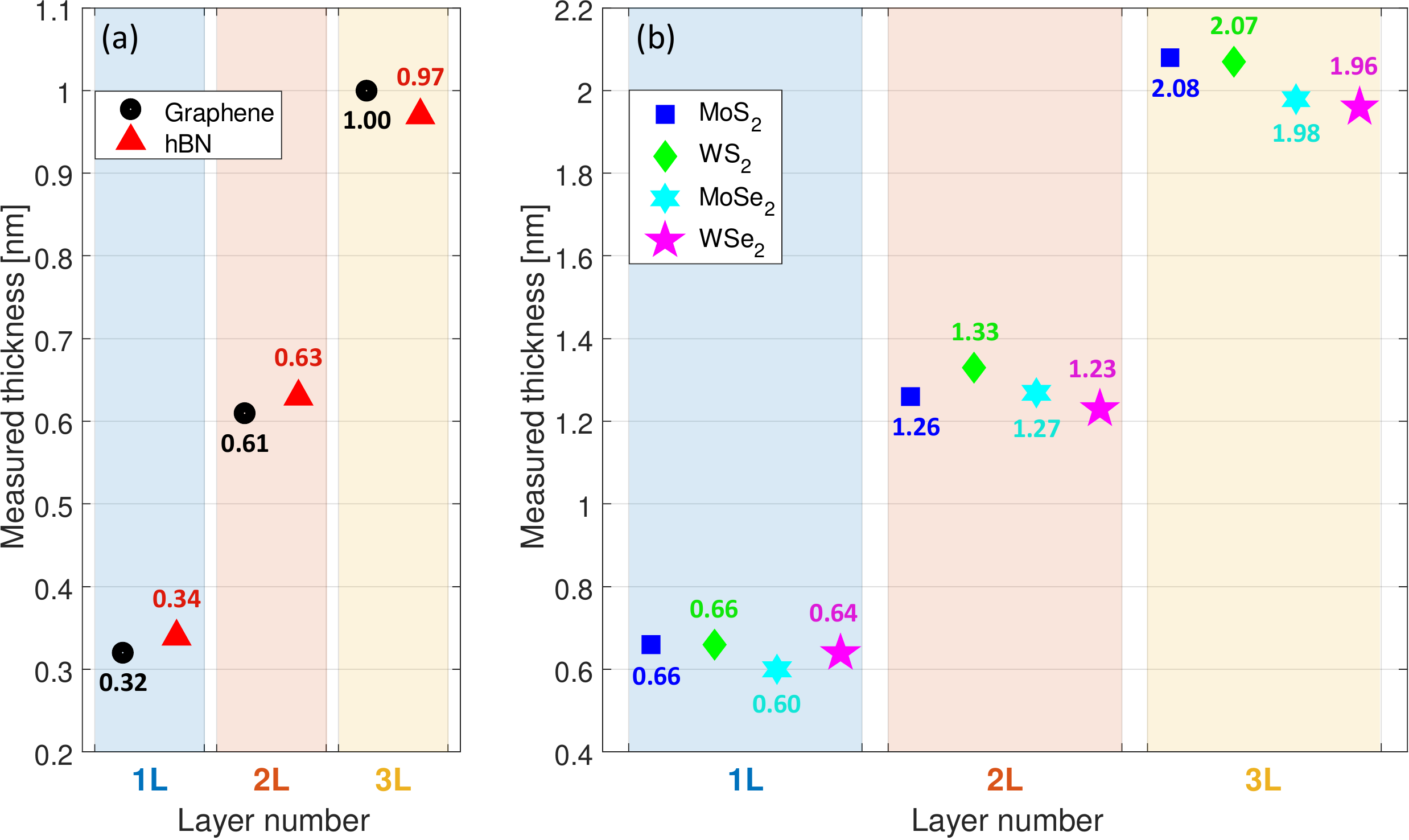}
     \caption{The SME derived thicknesses of mono-, bi- and trilayer flakes of (a) graphene, hBN, (b) MoS$_2$, WS$_2$, MoSe$_2$ and WSe$_2$.}
\label{fig:allthicknesses}
\end{figure}

The same procedure performed on graphene is repeated on mono-, bi- and trilayer candidates of hBN, MoS$_2$, WS$_2$, MoSe$_2$ and WSe$_2$ (the details of the TMD samples, measurements, and analysis are elaborated in section \ref{restofmats_S} of the SI). Figure \ref{fig:allthicknesses} plots a summary of the measured thicknesses for these samples, where the bi- and trilayer flakes are expected to be integer multiples of the monolayer \cite{Fang2017ThicknessWSe2}. As shown, very good agreements with the single-layer thickness of 0.32 nm for hBN \cite{Molaei2021ANitride} and individual values between 0.6-0.7 nm for TMDs (MoS$_2$ - 0.67 nm \cite{Qin2021-phaseLens, Benameur2011VisibilityNanolayers}, WS$_2$ - 0.65 nm \cite{Kim2015EngineeringLayers}, MoSe$_2$ - 0.7 nm \cite{He2016LayerJunctions} and WSe$_2$ - 0.67 nm \cite{Benameur2011VisibilityNanolayers, Sahin2013AnomalousWSe2}) are found for 1-layer, 2-layers and 3-layers, as in the graphene measurements. All flakes are also analyzed by Raman spectroscopy for their layer numbers, which show good agreement with the SME results (see SI). Among the flakes investigated, hBN holds special importance due to its transparency (especially its monolayer) under optical microscope and thus the entire process of its thickness measurement shall be detailed later in the section.

\begin{figure}[!b]
\centering
\includegraphics[width=\textwidth]{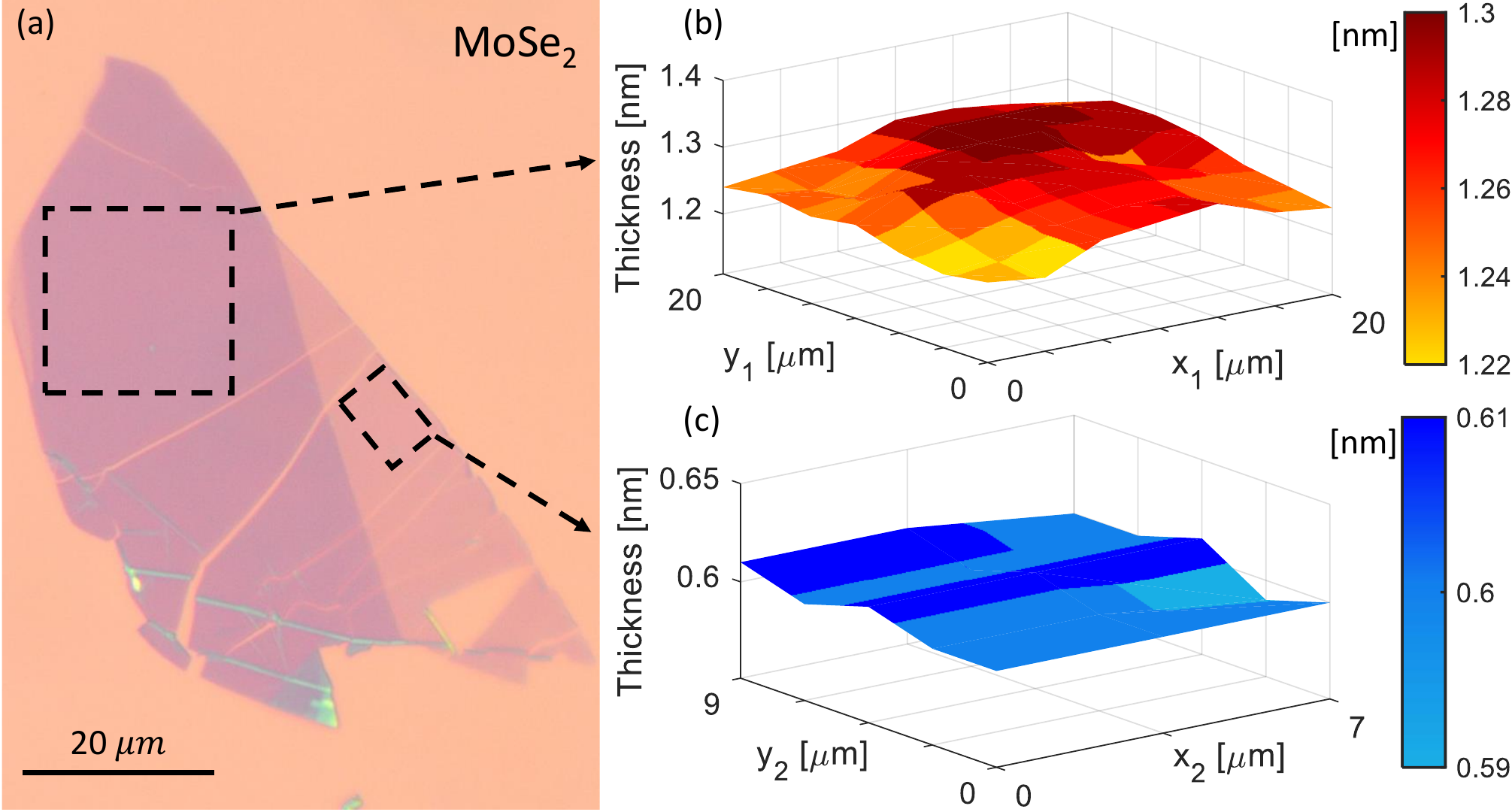}
     \caption{(a) Microscope image of an exfoliated MoSe$_2$ flake with marked areas on bilayer (square) and monolayer (rectangle) regions. The thickness mapping results by the SME of the (b) bilayer and (c) monolayer areas with respective (x$_1$, y$_1$) and (x$_2$, y$_2$) coordinates.}
\label{fig:scan}
\end{figure}

To showcase the reliability and sensitivity of the SME for mapping thickness variations of flakes, thickness mapping scans on monolayer and bilayer of MoSe$_2$ are performed. Figure \ref{fig:scan}(a) shows the optical microscope image of the MoSe$_2$ flake with monolayer and bilayer areas. The marked bilayer area of 20 $\mu m$ $\times$ 20 $\mu m$ is mapped with a spot size of 5 $\mu m$ and a step size of 2.5 $\mu m$ (49 points); and the monolayer area of 7 $\mu m$ $\times$ 9 $\mu m$ is mapped with a step size of 1 $\mu m$ (15 points). The local thickness variations in the bilayer and the monolayer scan measurements are plotted in Figure \ref{fig:scan}(b-c). The mean values of 1.266 nm and 0.603 nm with deviations of $\pm$0.04 nm and $\pm$0.01 nm are obtained in the scan measurements of bilayer and monolayer areas respectively. In order to understand the nature of these thickness deviations, repeatability measurements are performed 10 times on the same points in bilayer and monolayer areas to obtain the instrumental thickness accuracy, resulting in a deviation of $\pm$0.005 nm for both layers. An inference can be drawn that for both mono- and bilayer, the thickness variation originates from the flake's landscape as the deviation obtained in the scanning measurement is more than the instrumental accuracy.

\begin{figure}[!ht]
\centering
\includegraphics[width=\textwidth]{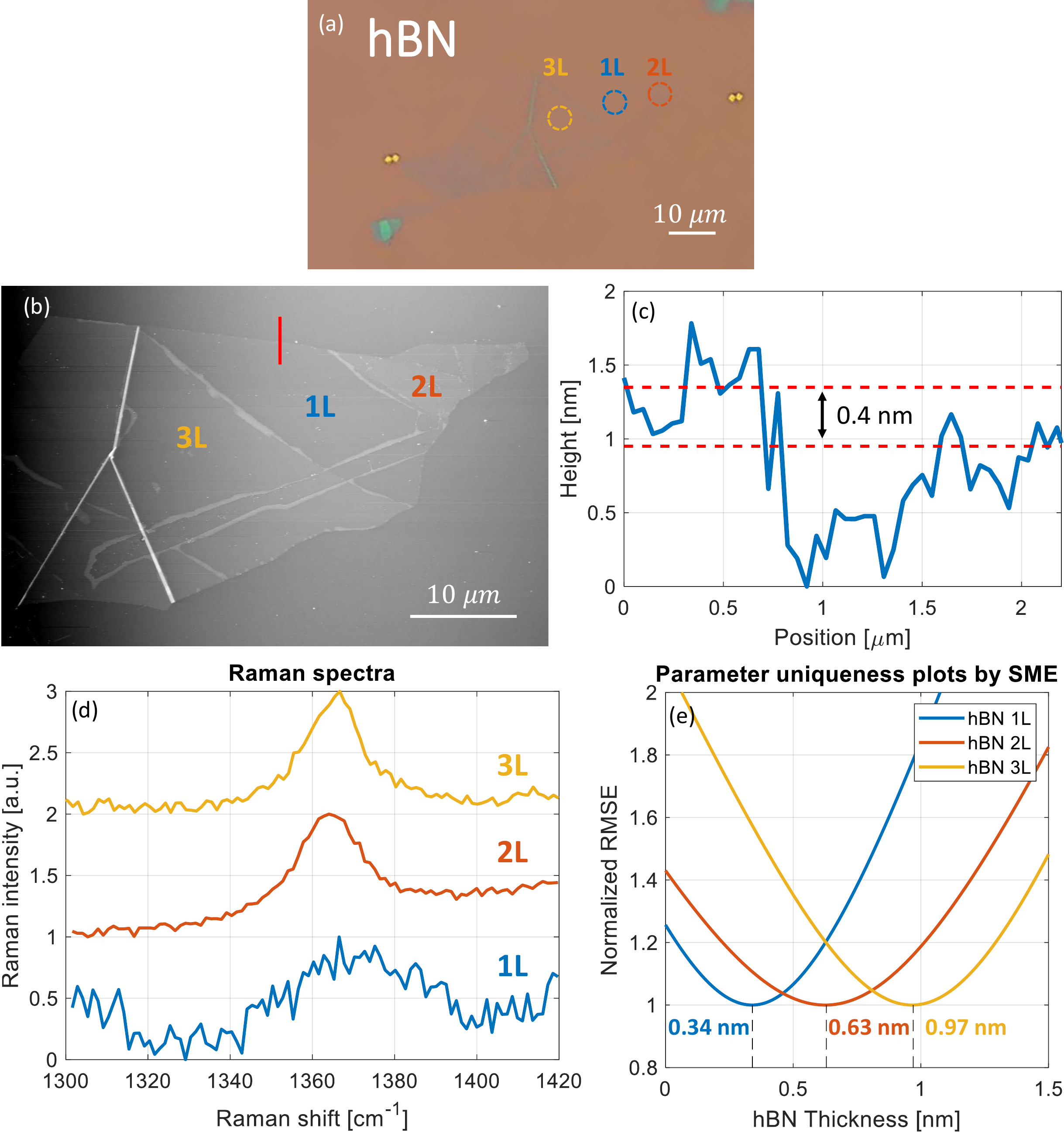}
     \caption{(a) Optical microscope image (contrast-enhanced for better visibility) of monolayer, bilayer and trilayer hBN with illustrated 5 $\mu m$ diameter SME measurement spots in blue, orange and yellow respectively. (b) AFM image of the hBN flake with the red line marking the transition of from the substrate to the monolayer hBN, and (c) the thickness profile of the red line showing a thickness of $\sim$0.4 nm, confirming the monolayer nature of the flake. (d) The measured Raman spectra of the mono-, bi- and trilayer hBN. (e) The parameter uniqueness plots by the SME pointing to thickness results of 0.34 nm, 0.63 nm and 0.97 nm for monolayer, bilayer and trilayer hBN respectively.}
\label{fig:hBN}
\end{figure}

Next, we show an advantage of the SME over currently used methods by performing thickness measurements on an hBN flake residing on a silicon substrate with 285 nm SiO$_2$, shown in Figure \ref{fig:hBN}. Incidentally the mono-, bi- and trilayers were found on a single flake at different locations as marked in the optical microscope image in Figure \ref{fig:hBN}(a). The optical contrast of the image has been amplified considerably using image processing tools to make the monolayer a bit more visible. However, it is to be noted that such tools are normally unavailable with a stand-alone optical microscope generally used for locating flakes, making the task rigorous. Even with such amplifications, the monolayer boundary is hardly discernible and only apparent in the AFM image of Figure \ref{fig:hBN}(b). The normalized and vertically displaced Raman spectra of the hBN flakes are plotted in Figure \ref{fig:hBN}(d). The relatively low-intensity, noisy peak centered at $\sim$1369 cm$^{-1}$ is the Raman signature for monolayer hBN \cite{Gorbachev2011HuntingSignatures}. Similar peak positions of bilayer and trilayer hBN between 1365-1366 cm$^{-1}$ were demonstrated in the literature \cite{Gorbachev2011HuntingSignatures}. These low signal and tiny spectral shifts compared to their spectral width make Raman spectra inconclusive in distinguishing between bi- and trilayers of hBN. In a similar manner, the AFM analysis performed on the monolayer hBN showed a thickness of $\sim$0.4 nm which is close to the reported values, as seen in Figure \ref{fig:hBN}(c). However, evidently the AFM height profile for the monolayer was noisy and less reliable. Comparatively, SME provides thickness results with much better confidence as inferred from the parameter uniqueness plot shown in Figure \ref{fig:hBN}(e). As seen in Figure \ref{fig:hBN}(e), the SME clearly distinguishes between mono-, bi- and trilayers of hBN with thickness results in agreement with integer multiples of monolayer thickness of 0.32 nm \cite{Molaei2021ANitride}. These results clearly demonstrate the superiority of the thickness measurements by the SME. 

Finally, we also prove the substrate-independent performance of our method with a number of measurements on graphene, WS$_2$ and hBN performed on silicon wafers with a different SiO$_2$ thickness of 90 nm, showing results consistent and as accurate to those discussed above (the results are detailed in section \ref{on90nm_S} of the SI).

\section{Summary and Conclusions}

A fast and accurate Fourier plane spectroscopic micro-ellipsometer is demonstrated for high resolution thickness mapping and thus layer number estimation of exfoliated vdW materials. The method allows a seamless combination of the ellipsometric measurement capabilities with any high resolution spectroscopic imaging setup or a microscope.

Six different types of vdW materials are measured and two different substrates are used to prove the sample- and substrate-independent performance of the proposed method. The SME could consistently identify among mono-, bi- and trilayers of the investigated materials with sub-angstrom precision. Especially, the SME could discreetly identify monolayer hBN on 285 nm Si/SiO$_2$ wafers, which is a challenging proposition for other characterization techniques. Repeatability measurements performed on various flakes exhibited minimal uncertainty in layer thicknesses, correctly identifying the layer numbers in the process. The high lateral resolution and the high accuracy are utilized to map thickness variations across flakes.

These results allow automated search of desired number of layers and mapping the thickness homogeneity of vdW materials. Such an automated system with an affordable and easily integrable accessory to an optical microscope might be highly coveted to the vdW community. Moreover, the SME can be easily used to extract optical constants from 2D flakes which opens up another domain for further research.

\section{Acknowledgements}

RR acknowledges support from the Israeli Science Foundation Grant 836/17 and from the NSF-BSF Grant 2019737. HS acknowledges funding from the Israeli Science Foundation Grant 861/19.

\clearpage
\bibliography{references}
\bibliographystyle{ieeetr}

%TC:ignore
\newpage
\supplementaryinformation

Here the spectroscopic micro-ellipsometer (SME) thickness measurements and the Raman spectra are elaborated for transition metal dichalcogenides (TMDs) of MoS$_2$, WS$_2$, MoSe$_2$, WSe$_2$.

\section{MoS$_2$, WS$_2$, MoSe$_2$ and WSe$_2$ results}
\label{restofmats_S}

\begin{figure}[!h]
\centering
\includegraphics[width=\textwidth]{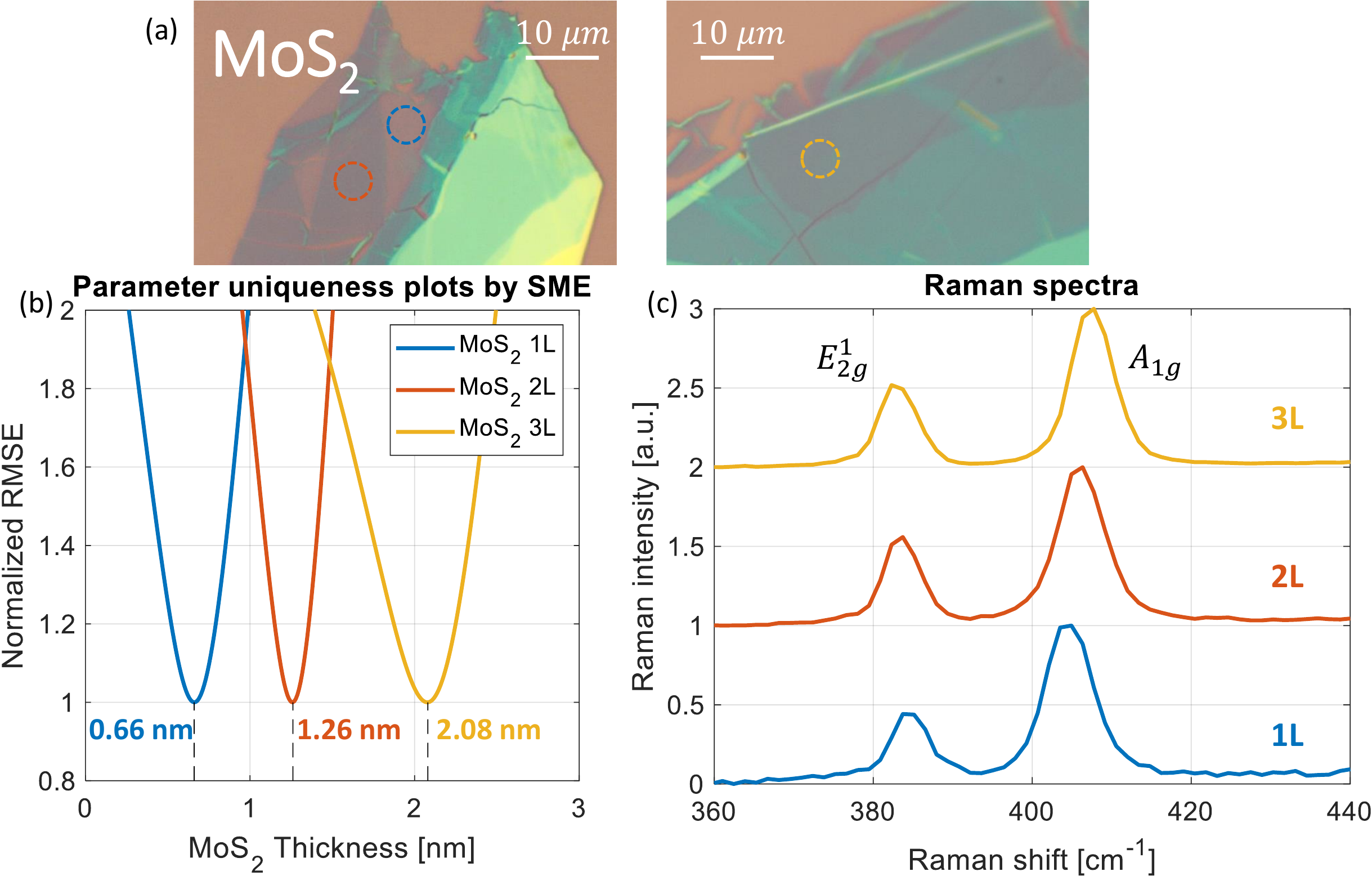}
     \caption{(a) Optical microscope images of monolayer, bilayer and trilayer MoS$_2$ with illustrated 5 $\mu m$ diameter SME measurement spots in blue, orange and yellow respectively. (b) The parameter uniqueness plots by the SME pointing to thickness results of 0.66 nm, 1.26 nm and 2.08 nm for monolayer, bilayer and trilayer MoS$_2$ respectively. (c) The measured Raman spectra of the same flakes.}
\label{fig:MoS2}
\end{figure}

\begin{figure}[!t]
\centering
\includegraphics[width=\textwidth]{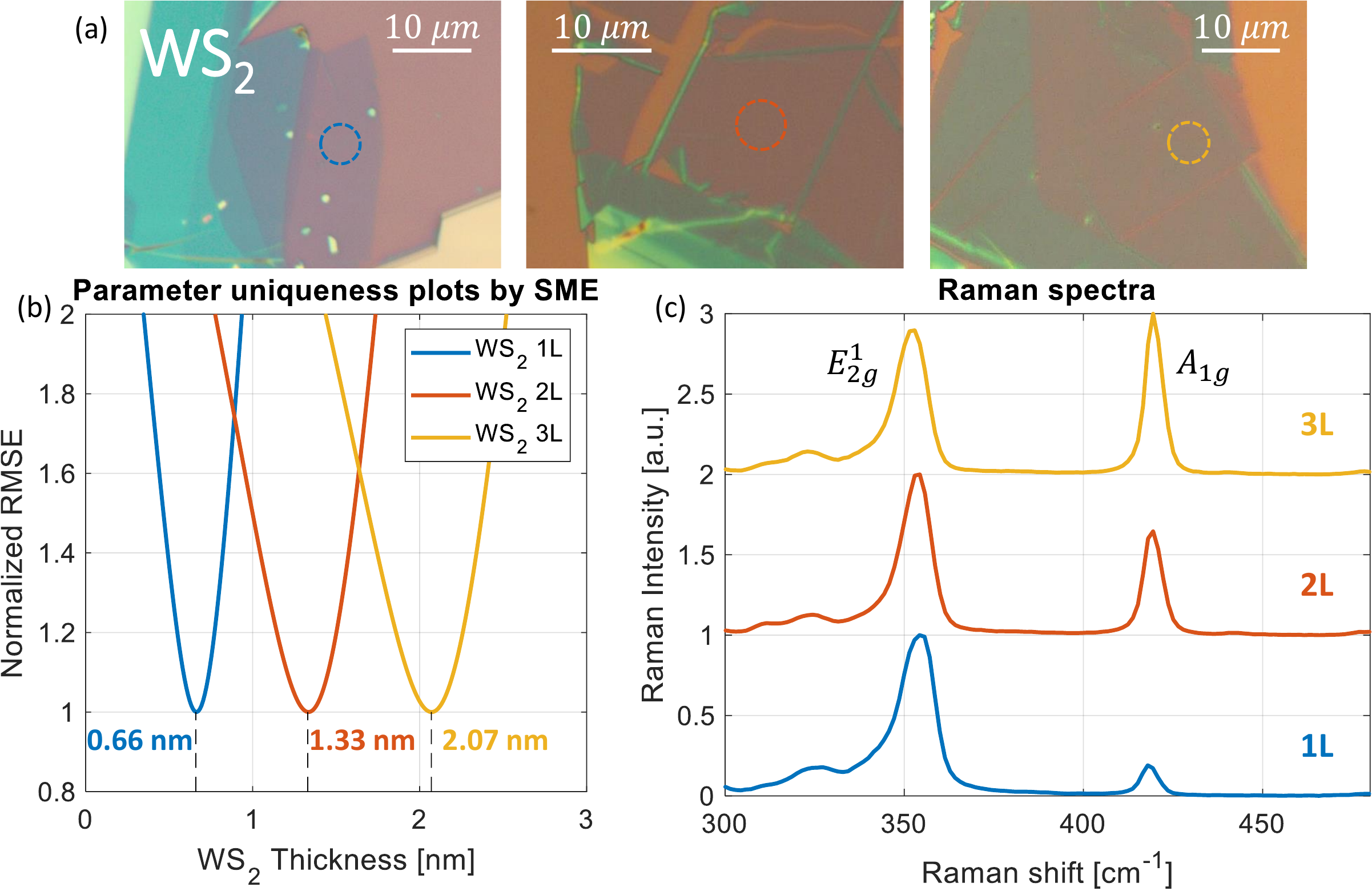}
     \caption{(a) Optical microscope images of monolayer, bilayer and trilayer WS$_2$ with illustrated 5 $\mu m$ diameter SME measurement spots in blue, orange and yellow respectively. (b) The parameter uniqueness plots by the SME pointing to thickness results of 0.66 nm, 1.33 nm and 2.07 nm for monolayer, bilayer and trilayer WS$_2$ respectively. (c) The measured Raman spectra of the same flakes.}
\label{fig:WS2}
\end{figure}

\begin{figure}[!h]
\centering
\includegraphics[width=\textwidth]{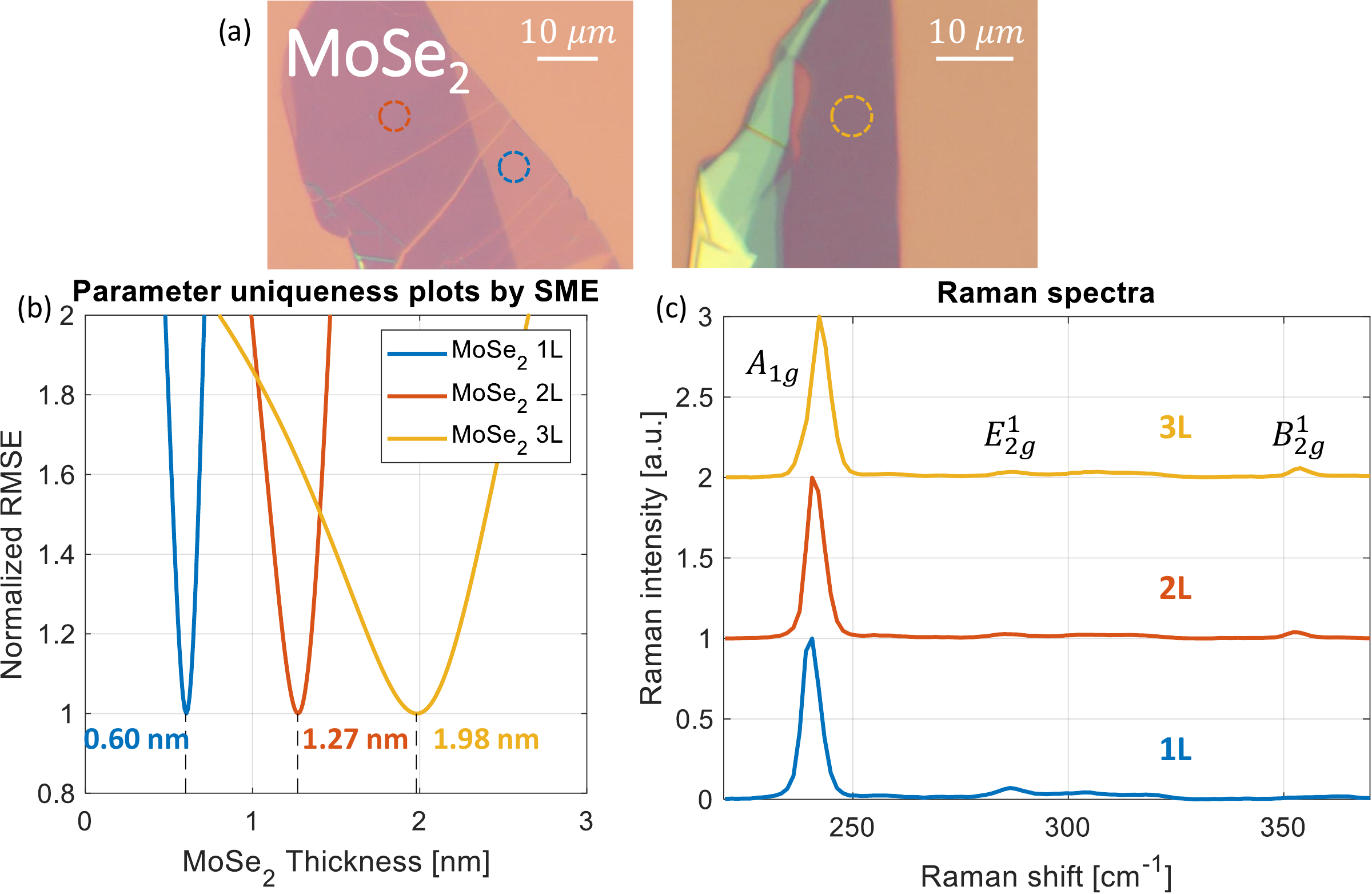}
     \caption{(a) Optical microscope images of monolayer, bilayer and trilayer MoSe$_2$ with illustrated 5 $\mu m$ diameter SME measurement spots in blue, orange and yellow respectively. (b) The parameter uniqueness plots by the SME pointing to thickness results of 0.60 nm, 1.27 nm and 1.98 nm for monolayer, bilayer and trilayer MoSe$_2$ respectively. (c) The measured Raman spectra of the same flakes.}
\label{fig:MoSe2}
\end{figure}

\begin{figure}[H]
\centering
\includegraphics[width=\textwidth]{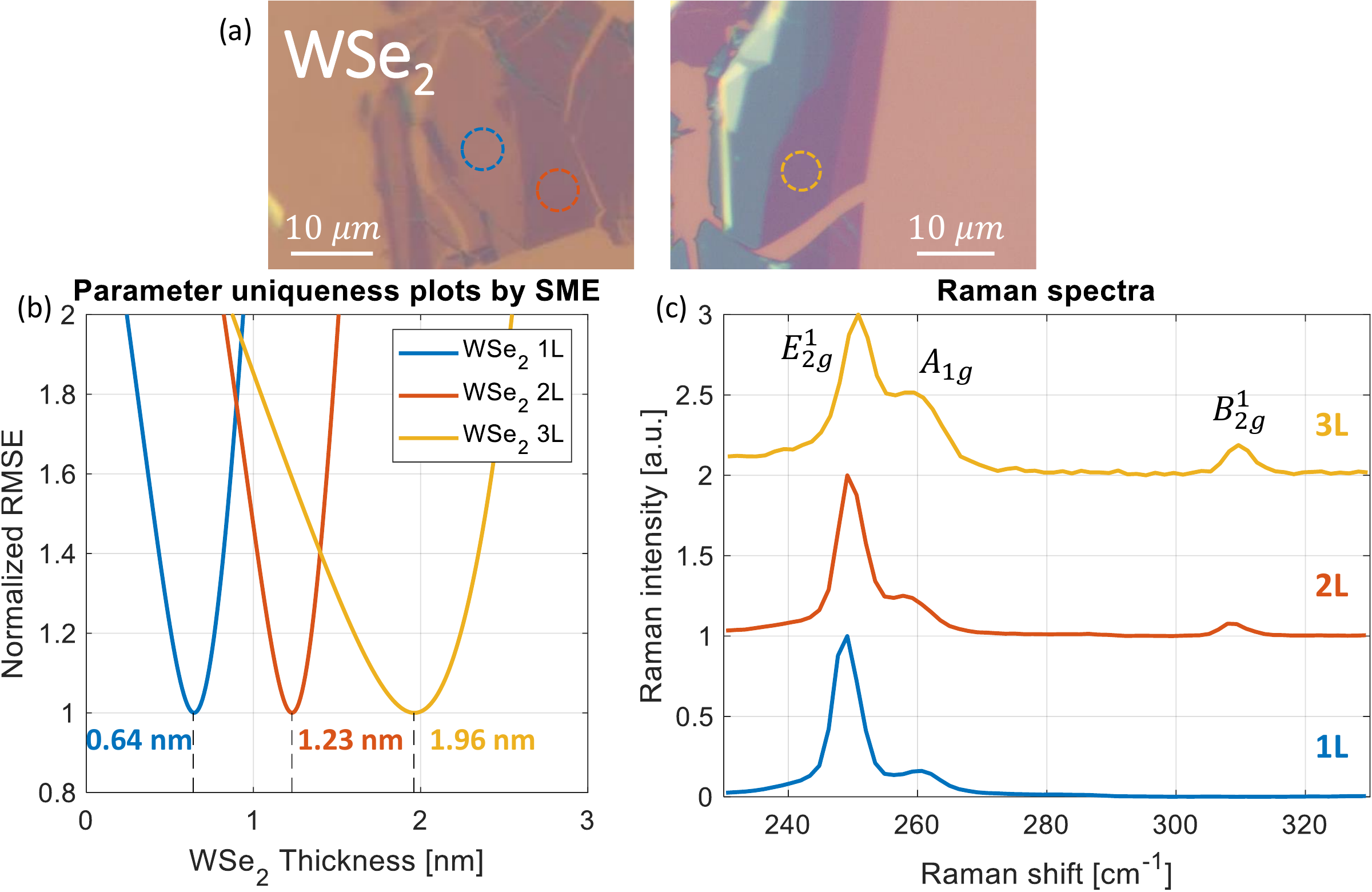}
     \caption{(a) Optical microscope images of monolayer, bilayer and trilayer WSe$_2$ with illustrated 5 $\mu m$ diameter SME measurement spots in blue, orange and yellow respectively. (b) The parameter uniqueness plots by the SME pointing to thickness results of 0.64 nm, 1.23 nm and 1.96 nm for monolayer, bilayer and trilayer WSe$_2$ respectively. (c) The measured Raman spectra of the same flakes.}
\label{fig:WSe2}
\end{figure}

Figures \ref{fig:MoS2}-\ref{fig:WSe2}(a) show the optical microscope images of mono-, bi- and trilayer flakes of MoS$_2$, WS$_2$, MoSe$_2$ and WSe$_2$, respectively. Figures \ref{fig:MoS2}-\ref{fig:WSe2}(b) are the parameter uniqueness plots by the SME from the ellipsometric measurements performed on the illustrated 5 $\mu m$ diameter areas on mono- (blue), bi- (orange) and trilayer (yellow) regions; resulting in thickness values of the measured flakes. All thickness results are within the tolerance limit from integer multiples of the monolayer thickness of 0.6-0.7 nm for the TMDs \cite{Qin2021-phaseLens, Benameur2011VisibilityNanolayers, Kim2015EngineeringLayers, He2016LayerJunctions, Sahin2013AnomalousWSe2}, as mentioned in the paper. Finally, Figures \ref{fig:MoS2}-\ref{fig:WSe2}(c) plot the normalized and vertically displaced (for clarity) measured Raman spectra of the flakes, where the excitation wavelength of the laser is 514.5 nm.

The measured Raman spectra of the MoS$_2$ flakes are plotted in Figure \ref{fig:MoS2}(c). The frequency difference ($\Delta\omega$) between the two Raman modes $E^1_{2g}$ and $A_{1g}$ for MoS$_2$ gives a strong indication for the number of layers. It has been shown that the $\Delta\omega$ for MoS$_2$ monolayer is $\sim$3 cm$^{-1}$ smaller than its bilayer, and the bilayer is $\sim$1.5 cm$^{-1}$ smaller than the trilayer \cite{Lee2010AnomalousMoS2}. This difference follows a decreasing trend until 6-layers and stabilizes at bulk state \cite{Lee2010AnomalousMoS2}. In Figure \ref{fig:MoS2}(c), the $\Delta\omega$ between the peaks are 19.8 cm$^{-1}$, 22.7 cm$^{-1}$ and 24.3 cm$^{-1}$ for mono-, bi- and trilayer respectively. This accounts to $\sim$3 cm$^{-1}$ and $\sim$1.5 cm$^{-1}$ difference between 1L-2L and 2L-3L respectively, confirming the mono-, bi- and trilayer nature of the MoS$_2$ flakes.

The measured Raman spectra of the WS$_2$ flakes are plotted in Figure \ref{fig:WS2}(c). As Raman fingerprints for the layer numbers of WS$_2$ flakes, the peak intensity ratios and frequency difference of the two peaks $E^1_{2g}$ and $A_{1g}$ at $\sim$350 cm$^{-1}$ and $\sim$420 cm$^{-1}$ respectively are taken into consideration. Peak intensity ratios of 4.5, 1.5 and 0.8 are received for the mono-, bi- and trilayer WS$_2$ measured in this work, respectively. A peak intensity ratio that is larger than 2 has been reported to be an exclusive signature for monolayer WS$_2$, followed by intensity ratios of roughly 1 and 0.7 for bilayer and trilayer respectively. In addition, the frequency difference between the peaks show an increasing trend as $\Delta\omega$ = 63.7 cm$^{-1}$, 65.8 cm$^{-1}$ and 67.2 cm$^{-1}$ for mono- to trilayer, which is in good agreement with the literature \cite{Berkdemir2013IdentificationSpectroscopy, Zeng2013OpticalDichalcogenides}.

The measured Raman spectra of the MoSe$_2$ flakes are plotted in Figure \ref{fig:MoSe2}(c). The out-of-plane $A_{1g}$ mode peak is found at 240.5 cm$^{-1}$ for monolayer MoSe$_2$, slightly shifting towards higher wavenumbers for bi-, and trilayers. The absolute intensity of this peak is highest for the bilayer (not shown here due to normalization). The lower wavenumber side of the $A_{1g}$ mode peak in the trilayer shows some broadening, probably due to existence of another vibrational component which is not resolved by the used Raman instrument. The weak in-plane $E^1_{2g}$ mode is found around 287 cm$^{-1}$ for monolayer and 286 cm$^{-1}$ for bilayer. The $B_{2g}$ mode at 353 cm$^{-1}$ does not exist for the monolayer and has the highest intensity for the bilayer. All these findings agree with the literature \cite{Tonndorf2013PhotoluminescenceWSe2} and confirms the monolayer, bilayer and trilayer nature of the flakes.

The measured Raman spectra of the WSe$_2$ flakes are plotted in Figure \ref{fig:WSe2}(c). The main vibrational mode $E^1_{2g}$ is located around 249 cm$^{-1}$ for the monolayer WSe$_2$, showing a slight blueshift with increasing number of layers. The intensity of this peak for the monolayer is the highest, being $\sim$2.5 times of the bilayer and $\sim$25 times of the trilayer (not shown here due to normalization). The $B^1_{2g}$ mode around 309 cm$^{-1}$ is not existent for the monolayer and shows the highest intensity counts for the bilayer. These Raman signatures concur with the literature \cite{Zeng2013OpticalDichalcogenides, Tonndorf2013PhotoluminescenceWSe2} and confirm that the measured flakes are indeed mono-, bi- and trilayers of WSe$_2$.

\section{Measurements on a different substrate}
\label{on90nm_S}

\begin{figure}[H]
\centering
\includegraphics[width=\textwidth]{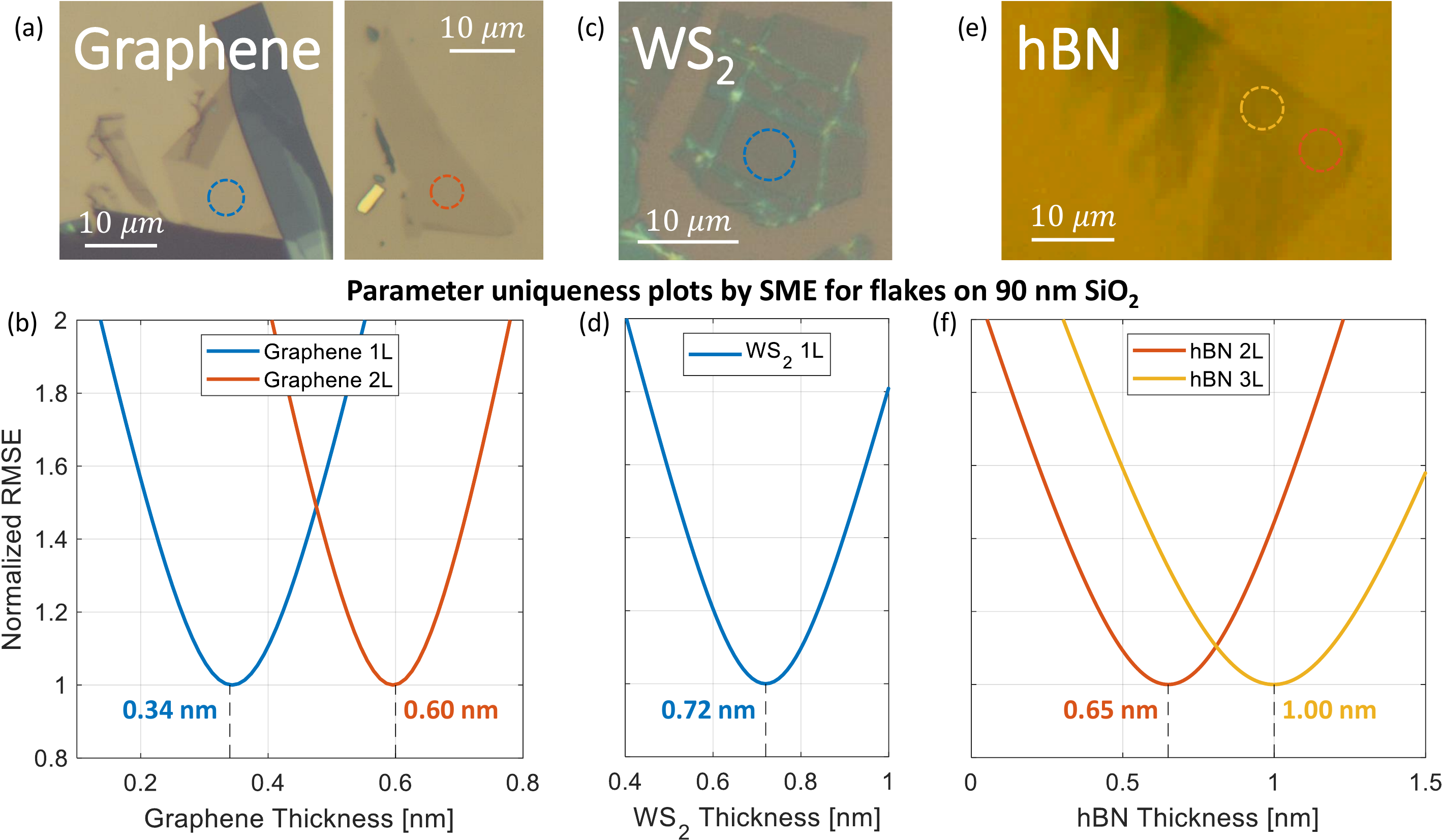}
     \caption{A number of additional flake measurements repeated on Si wafers with 90 nm of SiO$_2$ for the sake of demonstrating the sample-independent performance of the proposed method. (a) Graphene monolayer and bilayer optical microscope images and (b) their parameter uniqueness plots pointing to thickness results of 0.34 nm and 0.60 nm for the monolayer and bilayer respectively. (c) WS$_2$ monolayer optical microscope image and (d) its parameter uniqueness plot pointing to thickness result of 0.72 nm. (e) hBN bilayer and trilayer optical microscope images (contrast-enhanced for better visibility) and (f) their parameter uniqueness plots pointing to thickness results of 0.65 nm and 1.00 nm for the bilayer and trilayer respectively.}
\label{fig:90s}
\end{figure}

A number of measurements of various flakes are performed on substrates of silicon wafers with 90 nm SiO$_2$. Graphene, WS$_2$ and hBN are chosen to cover the whole range of materials discussed in the paper. Figure \ref{fig:90s} shows identical performance as obtained on silicon wafers with 285 nm SiO$_2$, proving the substrate-independent performance of the SME.

The Raman spectra of these flakes are also measured, resulting in very similar responses as demonstrated for the same materials on silicon wafers with 285 nm SiO$_2$.

%TC:endignore

\end{document}